\documentclass[a4paper,11pt]{iopart}
\usepackage{graphicx}
\usepackage{cite}
\usepackage{color}
\usepackage{amsfonts}
\usepackage{srcltx}
  \expandafter\let\csname equation*\endcsname\relax
  \expandafter\let\csname endequation*\endcsname\relax
\usepackage{amsmath}
\usepackage{amsthm}
\usepackage{mathtools}
\begin{document}
\title{On the upper bound of classical correlations in a bipartite quantum system}
\author{Theresa Christ and Haye Hinrichsen}
\address{Universit\"at W\"urzburg, Fakult\"at f\"ur Physik und Astronomie, Am Hubland, \\ 97074 W\"urzburg, Germany}

\ead{theresa.christ@physik.uni-wuerzburg.de, \\ \hspace{13mm}hinrichsen@physik.uni-wuerzburg.de}

\begin{abstract}
For a bipartite quantum system consisting of subsystems $A$ and $B$ it was shown by Zhang \textit{et al.}\ (Physics Letters A 376 (2012) 3588-3592) that the amount of classical correlations, which is used to define the quantum discord, is known to be bounded from above by the minimum of the von Neumann entropies of the subsystems $A$ and $B$. We provide an alternative proof that is shorter and more transparent as it works without defining correlation matrices.
\end{abstract}

\def\d{{\rm d}}
\def\0{\emptyset}
\def\ket#1{|#1\rangle}
\def\bra#1{\langle#1|}
\def\braket#1#2{\langle#1|#2\rangle}
\def\ketbra#1#2{|#1\rangle\langle #2|}
\newtheorem*{satz}{Theorem}
\newenvironment{beweis}{\begin{proof}}{\end{proof}}
\def\Tr{{\rm Tr}}

\newcommand{\GUE}{{\textsc{\tiny GUE}}}
\newcommand{\Haaralpha}{{\textsc{\scriptsize $\alphavec$}}}
\newcommand{\alphasub} {{\hspace*{-1pt}\textsc{\scriptsize $\alphavec$}}}
\newcommand{\Haarbeta}{{\textsc{\scriptsize $\vec\beta$}}}

\pagestyle{plain}
\noindent
In quantum information theory it is an important task to distinguish and quantify classical and quantum correlations in a mixed bipartite quantum system. It is well substantiated and accepted (see \cite{76.032327} and references therein) that the total amount of correlations is given by the quantum mutual information
\begin{equation}
 \mathcal{I}(A:B)=S(\rho_A)+S(\rho_B)-S(\rho_{AB})\,,
\end{equation}
where $\rho_{A(B)}=\Tr_{B(A)}[\rho_{AB}]$ is the reduced density matrix on subsystem $A(B)$ and $S(\cdot)$ denotes the von Neumann entropy $S(\rho)=-\Tr[\rho\log_2\rho]$. The total amount of correlations separates complementary into classical and quantum correlations (see \cite{76.032327} and references therein). The most widespread measure for quantum correlations is the entanglement of formation $E_F$ \cite{54.3824,entanglement}. However, the shortcoming of this measure is that it may be zero in situations where quantum correlations are present, as e.g.\ in Werner states \cite{88.017901}. Furthermore for certain states \cite{76.032327} it reaches higher values than the quantum mutual information $\mathcal{I}(A:B)$ does, which seems like entanglement of formation is no proper measure for quantum correlations \cite{76.032327}.\\
In 2001 Henderson and Vedral \cite{a13515} as well as Ollivier and Zurek \cite{88.017901} independently introduced an alternative measure for quantum correlations, called quantum discord. For a density matrix $\rho_{AB}$ on a bipartite Hilbert space $\mathcal{H}_{AB}=\mathcal{H}_A\otimes\mathcal{H}_B$ quantum discord is defined as the difference
\begin{equation}
 \mathcal{D}(A:B)=\mathcal{I}(A:B)-\mathcal{J}(A:B)
\end{equation}
between the total amount of correlations (classical plus quantum correlations) $\mathcal{I}(A:B)$ and the amount of classical correlations
\begin{equation}
\begin{split}
 &\mathcal{J}(A:B)=S(\rho_A)-\min\limits_{\{E_{B_k}\}}\sum\limits_k{p_k S(\rho_{A\mid k})}\\
 &\text{with } \rho_{A\mid k}=\frac{1}{p_k}\Tr_B[(\mathbf{1}_A\otimes E_{B_k})\rho_{AB}] \,,\quad p_k=\Tr[(\mathbf{1}_A\otimes E_{B_k})\rho_{AB}]\,.
\end{split}
\label{J}  
\end{equation}
Here $\{E_{B_k}\}$ denotes a positive operator valued measurement (POVM).\footnote{We follow the definition given in \cite{a13515}. The definition in \cite{88.017901} differs slightly, as they minimize over projective orthogonal measurements only.} 
Note that $\mathcal{I}(A:B)$ and $\mathcal{J}(A:B)$ coincide in classical systems whereas they generally differ in the quantum case \cite{105190502}. The main advantage of the quantum discord is that it responds properly to quantum correlations of separable states \cite{88.017901,81.052318}. Quantum discord is bounded from above by the marginal entropy of the measured subsystem \cite{zhang,0807.4490}, i.e.\ $\mathcal{D}(A:B)\leq S(\rho_B)$, while $\mathcal{D}(A:B)$ can reach higher values than $S(\rho_A)$ \cite{84.042124}. For further properties of the quantum discord see \cite{84.1655,1605.00806,1411.3208} and references therein. 
\\In this note we focus on the contribution of the classical correlations $\mathcal{J}(A:B)$ defined in eq.\ (\ref{J}). This quantity can be interpreted as the maximal classical information about subsystem $A$ that one can get by measuring $B$. Intuitively one would expect that this amount of correlation cannot exceed the information $S(\rho_A)$ contained in $A$. On the other hand, by measuring $B$ it is impossible to receive more information than $S(\rho_B)$. Together we therefore expect that \cite{82.052122}
\begin{equation}
 \mathcal{J}(A:B)\leq \min\bigl\{S(\rho_A),S(\rho_B)\bigr\}\,.
 \label{ineq}
\end{equation}
The inequality is sharp for pure states\cite{a13515}. Although this interpretation is quite intuitive, the inequality is non-trivial to prove. Zhang and Wu \cite{zhang} restricted the POVMs in definition (\ref{J}) to von Neumann measurements and provided a proof by defining a suitable quantum operation $\Psi(\rho_{AB})$ as well as the corresponding correlation matrix $\hat{\Psi}(\rho_{AB})$. They used a theorem by Roga \textit{et al.}\ \cite{105.040505}, which shows that the von Neumann entropy of the correlation matrix is an upper bound for the so-called Holevo quantity. Later Zhang \textit{et al.}\ \cite{zhang2} used the previous proof and generalized it to POVMs. 
\\In this paper we present an alternative proof for inequality (\ref{ineq}), which is based on two central theorems in quantum information theory, on Stinespring's and Neumark's theorem, and on the strong subadditivity of the von Neumann entropy. Our proof is shorter and also more transparent than the original one as it does not make use of correlation matrices.
\begin{satz}[Upper bound for classical correlations in a bipartite quantum system]
 Consider a bipartite quantum system with the density matrix $\rho_{AB}$ on a Hilbert space $\mathcal{H}_{AB}=\mathcal{H}_A\otimes\mathcal{H}_B$. The classical correlation measure $\mathcal{J}(A:B)$ given in equation (\ref{J}) is bounded from above by the von Neumann entropy of the subsystems $A$ and $B$, i.e.\ $\mathcal{J}(A:B)\leq \min\bigl\{S(\rho_A),S(\rho_B)\bigr\}$.
\end{satz}
\begin{beweis}
Thanks to Neumark's theorem\cite{peres} one can always write a POVM as an orthogonal projective measurement on an extended Hilbert space. Thus there exists an extension of $\mathcal{H}_{AB}$ to $\mathcal{H}_{AB\bar{B}}=\mathcal{H}_{AB}\otimes\mathcal{H}_{\bar{B}}$ so that the reduced state after the measurement is given~by
\begin{equation}
 \rho_{AB}'=\sum_k(\mathbf{1}_A\otimes E_{B_k})\rho_{AB}=\Tr_{\bar{B}}\big[\sum_k(\mathbf{1}_A\otimes \Pi_k)\rho_{AB\bar{B}}\big]\,,
\label{AB'}
\end{equation}
where $\rho_{AB\bar{B}}=\rho_{AB}\otimes\ketbra{\omega}\omega$ with an arbitrary normalized vector $\ket\omega \in \mathcal{H}_{\bar{B}}$ while $\{\Pi_k\}$ is a von Neumann measurement acting on $\mathcal{H}_{B\bar{B}}=\mathcal{H}_{B}\otimes\mathcal{H}_{\bar{B}}$. The reduced state after the measurement can also be expressed as $\rho_{AB}'=\Tr_{\bar{B}}[\rho_{AB\bar{B}}']$ with
\begin{equation}
 \rho_{AB\bar{B}}'=\sum_k(\mathbf{1}_A\otimes \Pi_k)\rho_{AB\bar{B}}(\mathbf{1}_A\otimes \Pi_k)\,.
\end{equation}
Inspired by \cite{0807.4490} we use Stinespring's theorem and write the projective orthogonal measurement $\{\Pi_k\}$ on subsystem $B\bar{B}$ as a unitary transformation $U$ on an extended Hilbert space $\mathcal{H}_{B\bar{B}}\otimes\mathcal{H}_C$:
\begin{equation}
\begin{split}
 \rho_{AB\bar{B}C}'&=\big(\mathbf{1}_A\otimes U\big)\big(\rho_{AB\bar{B}}\otimes\ketbra 0 0_C\big)\big(\mathbf{1}_A\otimes U^\dagger\big) \\
 &= \sum_{k,j}\big[\big(\mathbf{1}_A\otimes\Pi_k\big)\rho_{AB\bar{B}}\big(\mathbf{1}_A\otimes\Pi_j\big)\big]\otimes \ketbra{c_k}{c_j}
\end{split}
\end{equation}
with orthonormal vectors $\ket{c_k}\in \mathcal{H}_C$. Neumark's theorem and eq.\ (\ref{J}) give
\begin{equation}
\begin{split}
 p_k&=\Tr_{AB}[(\mathbf{1}_A\otimes E_{B_k})\rho_{AB}]=\Tr_{AB\bar{B}}[(\mathbf{1}_A\otimes \Pi_k)\rho_{AB\bar{B}}]\quad\text{and}\\
 \rho_{A\mid k}&=\frac{1}{p_k}\Tr_{B}[(\mathbf{1}_A\otimes E_{B_k})\rho_{AB}]=\frac{1}{p_k}\Tr_{B\bar{B}}[(\mathbf{1}_A\otimes \Pi_k)\rho_{AB\bar{B}}]\,,
\end{split}
\end{equation}
so that it is now straightforward to compute the reduced density matrices
\begin{equation}
\begin{split}
 \rho_{AC}'&=\Tr_{B\bar{B}}\bigl[\rho_{AB\bar{B}C}'\bigr]=\sum_{k}p_k \rho_{A\mid k}\otimes\ketbra {c_k}{c_k}\,,\\
 \rho_A'&=\Tr_C\bigl[\rho_{AC}'\bigr]=\sum_{k}p_k \rho_{A\mid k}=\rho_A\,,\\
 \rho_{B\bar{B}C}'&=\Tr_A\bigl[\rho_{AB\bar{B}C}'\bigr]=\sum_{j,k}\big(\Pi_k\rho_{B\bar{B}}\Pi_j\big)\otimes\ketbra{c_k}{c_j}=V\rho_{B\bar{B}} V^\dagger\,,\\
 \rho_{B\bar{B}}'&=\Tr_C\bigl[\rho_{B\bar{B}C}'\bigr]=\sum_{k}\Pi_k\rho_{B\bar{B}}\Pi_k=\sum_k p_k\Pi_k\,,
\end{split}
\end{equation}
where $V$ denotes a unitary mapping $V\colon \mathcal{H}_{B\bar{B}} \to \mathcal{H}_{B\bar{B}C}$ with $\ket v \mapsto \sum_k \Pi_k\ket v \otimes \ket{c_k}$, i.e.\ $\rho_{B\bar{B}}$ and $\rho_{B\bar{B}C}'$ are unitary equivalent. This yields the von Neumann entropies
\begin{equation}
\begin{split}
 S(\rho_{AC}')&=-\Tr\Big[\sum_{k}\big(p_k \rho_{A\mid k}\otimes\ketbra {c_k}{c_k}\big)\log_2\big(\sum_{j}p_j \rho_{A\mid j}\otimes\ketbra {c_j}{c_j}\big)\Big]\\
	      &=-\Tr\Big[\sum_{k}p_k \rho_{A\mid k}\log_2(p_k \rho_{A\mid k})\Big]=H(\{p_k\})+\sum_k p_k S(\rho_{A\mid k})\,,\\
 S(\rho_A')&=S(\rho_A)\,,\qquad S(\rho_{B\bar{B}C}')=S(\rho_{B\bar{B}})\,,\qquad S(\rho_{B\bar{B}}')=H(\{p_k\})
\end{split}
\end{equation}
where $H(\{p_k\})$ denotes the Shannon entropy of the probability distribution $\{p_k\}$. Applying the strong subadditivity in the alternative form 
\begin{equation}
  S(\rho_A')+S(\rho_{B\bar{B}}')\leq S(\rho_{AC}')+S(\rho_{B\bar{B}C}')
\end{equation}
given in \cite{105.040505} we obtain
\begin{equation}
  S(\rho_A)-\sum_k p_k S(\rho_{A\mid k})\leq S(\rho_{B\bar{B}})\,.
\end{equation}
As $S(\rho_{B\bar{B}})=S(\rho_B\otimes\ketbra{\omega}{\omega})=S(\rho_B)$ we finally get 
\begin{equation}
 \mathcal{J}(A:B)\leq S(\rho_B)\,.
\end{equation}
On the other hand $S(\rho_{A\mid k})\geq 0$ implies the additional inequality
\begin{equation}
 \mathcal{J}(A:B)\leq S(\rho_A)\,,
\end{equation} 
which completes the proof.
\end{beweis}
\vspace{5mm}
\noindent
\textbf{Acknowledgments}\\
We would like to thank P. Fries for useful discussions.

\section*{References}

\bibliography{literatur}
\bibliographystyle{unsrt}


\end{document}